\documentclass{jpp}
\usepackage{graphicx}
\usepackage{amsmath}	
\usepackage{amssymb}	
\usepackage{multicol}        
\usepackage{pdflscape}	
\usepackage{gensymb}
\usepackage{tikz}
\usetikzlibrary{arrows}

\newcommand{\beq}{\begin{equation}}
\newcommand{\eeq}{\end{equation}}

\newcommand{\vA}{v_{\rm A}}

\newcommand{\vu}{\mathbf{u}}
\newcommand{\vb}{\mathbf{b}}

\newcommand{\vz}{\mathbf{z}}

\newcommand{\vup}{\vu_\perp}
\newcommand{\vbp}{\vb_\perp}
\newcommand{\vzp}{\vz_\perp}

\newcommand{\lpar}{l_\parallel}

\newcommand{\kpar}{k_\parallel}

\newcommand{\tc}{\tau_{\rm c}}
\newcommand{\tclam}{\tau_{\mathrm{c}\lambda}}

\newcommand{\tck}{\tau_{\mathrm{c}k_\perp}}
\newcommand{\kp}{{k_\perp}}
\newcommand{\tcr}{\tau_{\mathrm{c}\rho}}

\shorttitle{Intermittency and dissipation in collisionless plasma turbulence}

\shortauthor{A. Mallet and others}

\title{Interplay between intermittency and dissipation in collisionless plasma turbulence}

\author{Alfred Mallet\aff{1}
	\corresp{\email{alfred.mallet@berkeley.edu}},
Kristopher G. Klein\aff{2,3},
Benjamin D. G. Chandran\aff{4},
Daniel Gro{\v s}elj\aff{5},
Ian  W. Hoppock\aff{4},
Trevor A. Bowen\aff{1,6},
Chadi S. Salem\aff{1} \and
Stuart D. Bale\aff{1,6}}
\date{\today}
\affiliation{\aff{4}Space Science Center, University of New Hampshire, Durham, NH 03824, USA
\aff{1}Space Sciences Laboratory, University of California, Berkeley CA 94720, USA
\aff{2}Climate and Space Sciences and Engineering, University of Michigan, Ann Arbor, MI 48109, USA
\aff{3}Lunar and Planetary Laboratory, University of Arizona, Tucson, AZ 85719, USA
\aff{5}Max-Planck-Institut f\"ur Plasmaphysik, D-85748 Garching, Germany
\aff{6}Physics Department, University of California, Berkeley CA 94720, USA}
\begin{document}
\maketitle
\begin{abstract}
We study the damping of collisionless Alfv\'enic turbulence in a strongly magnetized plasma by two mechanisms: stochastic heating (whose efficiency depends on the local turbulence amplitude $\delta z_\lambda$) and linear Landau damping (whose efficiency is independent of $\delta z_\lambda$), describing in detail how they affect and are affected by intermittency. The overall efficiency of linear Landau damping is not affected by intermittency in critically balanced turbulence, while stochastic heating is much more efficient in the presence of intermittent turbulence. Moreover, stochastic heating leads to a drop in the scale-dependent kurtosis over a narrow range of scales around the ion gyroscale.
\end{abstract}
\section{Introduction}
The question of how collisionless plasma turbulence dissipates via kinetic processes has received a great deal of recent interest \citep{parashar2015}. The heating mechanism(s) that effect this dissipation have dramatic consequences for the basic thermodynamic state of the plasma, controlling the ion-to-electron temperature ratio as well as affecting the temperature anisotropy of the plasma with respect to the local magnetic-field direction.
Attempts at solving this problem often fall into one of two camps: (i) studies that invoke the ``quasilinear premise" \citep{klein2012,howes2014} and propose that turbulent fluctuations damp at the same rate (e.g., the linear Landau damping rate \citep{landau1946}) as linear plasma waves with similar polarization properties \citep{howes2006,howes2008,schektome2009,howesweakened2011,tenbarge2013,tenbarge2013b,told2015,howes2018}, or, alternatively, (ii) studies that focus on intermittency and the associated ``coherent structures" \citep{burlaga1991,horburyinterm,sorrisovalvo1999,bruno2007,salem2009,greco09,cho2009,parashar2009,greco2010,osman2011,osman2012a,osman2012b,perri2012,greco2012,wu2013,karimabadi2013,osman14rec,chasapis15,lion2016,perrone2016,navarro2016,wan2016,matthaeuslamkin86,servidio2009,servidio2011}, 
arguing that these structures dissipate in a fundamentally different way than linear waves.
In this Letter, we straddle these two camps by developing a novel modelling framework for the damping of intermittent turbulence. We use this to predict, for the first time, the quantitative dependence (or independence) of different heating mechanisms on the level of intermittency, and the effect (or lack of effect) of these heating mechanisms on the intermittency itself, with several surprising results. These results suggest a simple new observational test, based on the scale-dependent kurtosis near the ion gyroscale, that will allow us to distinguish between different heating mechanisms in collisionless plasma turbulence, for example in the solar wind.

We first show that, in intermittent, critically balanced turbulence, intermittency has no effect upon the total turbulent heating rate resulting from linear Landau damping, and that linear Landau damping has no effect on the level of intermittency. These results only apply when the turbulence is critically balanced: in both weak turbulence and (unphysical) isotropic turbulence, the linear Landau heating rate does depend on the intermittency. Thus, the ``linear" nature of Landau damping does not by itself make (as might na\"ively be expected) its associated turbulent heating rate independent of intermittency. We then contrast this with the mechanism of \emph{stochastic heating} \citep{mcchesney1987,chen2001,white2002,voitenko2004,chaston2004,fiksel2009,chandran2010,chandran2010b,chandran2011,bourouaine2013,vech2017}; when the turbulence amplitude at the ion gyroscale $\rho=v_{\rm th}/\Omega_i$ (where $v_{\rm th}=\sqrt{2T_i/m_i}$ is the ion thermal speed and $\Omega_i=ZeB/m_i$ is the ion gyrofrequency) becomes sufficiently large, ion orbits become chaotic, and ions may gain energy by interacting with gyroscale turbulent structures with frequencies much less than $\Omega_i$.
 The stochastic damping rate is a highly nonlinear function of turbulent fluctuation amplitude.
We show that (i) intermittency dramatically increases the overall stochastic heating rate, and (ii) stochastic heating reduces the scale-dependent kurtosis of the turbulent fluctuations at the scale $\rho$. Finally, we show that because of this strong dependence on intermittency, stochastic heating may remain an important dissipation mechanism in astrophysical situations, where na\"ively it would be ignored due to the small overall turbulence amplitude at the gyroscale.

\section{Intermittency model}
We restrict our analysis to intermittent Alfv\'en-wave turbulence and damping mechanisms that are effective at $\kp\rho \lesssim 1$. We assume that the velocity and magnetic field fluctuations (in velocity units) are much smaller than the background magnetic field, and that the fluctuations are highly anisotropic with respect to the direction of the background magnetic field, i.e. their parallel wavevectors are much smaller than their perpendicular wavevectors, $k_\parallel \ll k_\perp$. 
This allows us to model the turbulence with the equations of reduced magnetohydrodynamics (RMHD) \citep{strauss1976,kadomtsev1973,montgomery1982}, compactly written in terms of \citet{elsasser} variables $\vzp^\pm = \vup\pm\vbp$, where $\vup$ and $\vbp$ are the perpendicular velocity and magnetic-field (in velocity units) fluctuations respectively. There are a number of different intermittency models \citep{mullerinterm2000,Chandran14} available
; here, we will use the MS17 \citep{ms16} model, but our results do not depend in detail on this choice \footnote{Provided that the critical balance conjecture \citep{gs95} is incorporated in an appropriate way -- this will be explained further in Section \ref{sec:lld}.}. The Elsasser fluctuation amplitude of a structure with perpendicular scale $\lambda$ is a random variable,
\beq
\delta z_{\lambda} = \delta z_{L_\perp} \Delta^q,\label{eq:logpois}
\eeq
where $\delta z_{L_\perp}$ is the outer scale amplitude, the constant $\Delta = 1/\sqrt{2}$, and $q$ is a Poisson random variable \footnote{This statement is a slight simplification: in fact, it is a weighted combination of Poissons, which nevertheless exhibits the same statistics.}
 with mean $\mu=-\log{(\lambda/L_\perp)}$, $L_\perp$ being the outer scale. This distribution has ``heavy tails", becoming heavier at smaller scales $\lambda$, a classic hallmark of intermittency \citep{frischbook}. This may be usefully quantified by the scale-dependent kurtosis,
 \beq
\kappa_{\lambda} \equiv \frac{\langle\delta z_\lambda^4\rangle}{\langle\delta z_\lambda^2\rangle^2}=\kappa_{L_\perp} \left(\frac{\lambda}{L_\perp}\right)^{-1/4}.\label{eq:kurt}
\eeq
The nonlinear and linear timescales of each structure are
\begin{align}
\tau_{\mathrm{nl}\lambda} \sim \frac{\lambda}{\delta z_\lambda \sin\theta_\lambda},\quad
\tau_{\mathrm{A}\lambda} \sim \frac{l_{\parallel\lambda}}{\vA}\label{eq:times}
\end{align}
respectively, where $\vA=B_0/\sqrt{4\pi n_i m_i}$ is the Alfv\'en speed, $\theta_\lambda$ is the ``alignment angle" \citep{boldyrev}, and
\beq
\sin\theta_\lambda \sim \left(\frac{\lambda}{L_\perp}\right)^{1/2}\frac{\delta z_{L_\perp}}{\delta z_{\lambda}}.\label{eq:theta}
\eeq
This model incorporates \emph{refined critical balance} \citep{rcb}: The linear and nonlinear timescales in each structure are comparable, $\chi_\lambda \equiv \tau_{\mathrm{A}\lambda}/\tau_{\mathrm{nl}\lambda} \sim 1$\footnote{This principle is obeyed both in numerical simulations of RMHD turbulence \citep{rcb}, and in the solar-wind turbulence \citep{chen2016}.}. Thus, either time may be used as the cascade timescale $\tclam$. The cascade power within the local subvolume of a particular structure is
\beq
\epsilon_\lambda \sim \frac{\delta z_\lambda^2}{\tclam}.\label{eq:eflux}
\eeq
Note that 
$
\langle \epsilon_{\lambda}\rangle = {\delta z_{L_\perp}^3}/{L_\perp}\equiv \epsilon
$,
the injected power, for $\lambda$ in the inertial range. 
\section{Damping model}
In this work, we will assume that the damping mechanisms we study irreversibly dissipate energy that is removed from the Alfv\'enic cascade \footnote{Technically, this irreversibility arises due to collisions \citep{schektome2009,zocco2011,loureiro2013,navarro2016,pezzi2016,servidio2017}. 
There is certainly no guarantee that the collisional heating occurs in the same spatial location as the damping: for example, \citet{navarro2016} found that in gyrokinetic turbulence, collisional heating was not localized to current sheets, because it takes some time for the velocity-space features generated near the current sheet to phase-mix to small velocity-space scales, and during this time, the plasma containing those velocity-space features flows away from the current sheet \citep{schektome2009}.
Where exactly the final collisional entropy production happens does not affect the results in this paper.}. 
We can then relate the heating rate $Q_\lambda$ to the damping rate $\gamma_\lambda$ via
\beq
Q_{\lambda} \sim \gamma_\lambda \delta z_\lambda^2.
\eeq

To motivate our model, we begin with the nonintermittent cascade model of Howes \emph{et al.} \citep{howes2008,howesweakened2011,batchelor1953}, which
in steady state far from the forcing wavenumber leads to
\beq
\frac{\epsilon_{k_1}}{\epsilon_{k_0}} = \exp\left(-\int_{k_0}^{k_1} 2\gamma_{k_\perp}\tck \frac{d \kp}{\kp}\right),
\eeq
where $\epsilon_{k_\perp}={(\delta z_{k_\perp})^2}/{\tau_{\mathrm{c}k_\perp}}\label{eq:epsnoint}$
is the cascade power at perpendicular wavenumber $k_\perp$, $\delta z_{\kp}$ is the turbulence amplitude at $\kp$, and $\gamma_{\kp}$ is the damping rate at $\kp$. 
In order to investigate different damping mechanisms analytically, we make the simplifying assumption that the damping is localised to one particular reference scale $\rho$, i.e. 
$
\gamma_\kp\tck = \gamma_\rho\tau_{\mathrm{c}\rho} \delta[\log(\kp\rho)],
$
where $\delta[\ldots]$ denotes the Dirac delta distribution. In practice, various potentially important forms of damping are localised around the ion gyroscale: for example, stochastic heating, and ion Landau damping at high $\beta_i$ \citep{howes2006}. The cascade power ($\epsilon_{\rho-}$) and turbulence amplitude ($\delta z_{\rho-}$) at $\kp\rho=1+\mathrm{d}$ may then be written in terms of their counterparts $\epsilon_{\rho+}$ and $\delta z_{\rho+}$ at $\kp\rho=1-\mathrm{d}$ (where $\mathrm{d}\ll1$):
\begin{align}
\epsilon_{\rho-} &= \epsilon_{\rho+}\exp\left(-2\gamma_\rho\tau_{\mathrm{c}\rho}\right),
\label{eq:damping_eps}
\\
\delta z_{\rho-} &= \delta z_{\rho+}\exp\left(-\frac{2}{3}\gamma_\rho\tau_{\mathrm{c}\rho}\right),
\label{eq:masterdamp}
\end{align}
where to obtain Eq.~(\ref{eq:masterdamp}) we use Eq.~(\ref{eq:times}), assuming that damping affects the amplitude but not the dynamic alignment \footnote{Physically, this formula seems reasonable: Exponential damping of a structure's amplitude $\delta z_{\rho+}$ at rate $2\gamma_\rho/3$ over its lifetime $\tau_{\mathrm{c}\rho}$ leads to precisely Eq.~(\ref{eq:masterdamp}). The RMHD timescales [Eq.~(\ref{eq:times})] do not technically apply beyond the ion gyroscale; we assume the true dynamical timescales are continuous in $\lambda$ (in the absence of damping), allowing us to use Eq.~(\ref{eq:times}) to derive Eq.~(\ref{eq:masterdamp}).}.

To generalise this, note that if a turbulent structure has perpendicular scale $\lambda$ and amplitude $\delta z_\lambda$, its fluctuation power $\delta z_{\kp}\sim\delta z_\lambda$ peaks at $\kp \sim 1/\lambda$.  
We further assume \citep{k62} that the local values of random variables in a structure set its dynamical timescales $\tclam$, $\gamma_{\lambda}^{-1}$, and promote all the variables in Eqs.~(\ref{eq:damping_eps}-\ref{eq:masterdamp}) to configuration-space random variables. We call $\gamma_\rho\tcr$ the \emph{damping factor}. 

We would like to stress that, of course, collisionless damping mechanisms do not appear in RMHD, which models the (undamped) Alfv\'enic fluctuations at $k_\perp\rho_i \ll 1$ (irrespective of collisionality). However, the intermittency at $k_\perp\rho_i \sim 1$, where collisionless damping appears in more complete models, is almost entirely produced by the turbulence in the (assumed to be long, $L_\perp/\rho_i \gg 1$) inertial range in which RMHD is a good approximation. Thus, we model the intermittency using RMHD, and then add the dissipation in the simple way described above at the scale at which the RMHD approximation begins to break down.

We have made the rather drastic simplification that the damping only occurs over an infinitesimal scale interval. No real damping mechanism is truly this localized in scale. To go beyond this approximation, one would have to simultaneously integrate over scale not only the damping part of the process (as in \citealt{howes2008}) but also the random part of the evolution describing the random log-Poisson evolution of the intermittent probability distribution of amplitudes; replacing the algebraic exponents of Eqs.~(\ref{eq:damping_eps}) and (\ref{eq:masterdamp}) with functional integrals. This makes the model analytically intractable. Moroever, across any particular individual scale, the incremental damping of the fluctuations is well described by Eqs.~(\ref{eq:damping_eps}) and (\ref{eq:masterdamp}), which means that many of our results will not be qualititatively altered by making this approximation.

Finally, it is worth mentioning that other timescales could potentially enter the problem \citep{matthaeus2014}; for example, waves could be excited via instability of the particle velocity distribution function. Indeed, \citealt{klein2018} have found that the majority of solar-wind plasma is unstable, although only about 10\% appears to be strongly unstable in that the growth time is shorter than their estimate of the cascade time. We have ignored this possibility in our analysis here, and assume that the underlying velocity distribution function is stable.
\section{Linear Landau damping}\label{sec:lld}
One important and well-studied damping mechanism is linear Landau damping \citep{howes2006}, for which the damping rate may be written (in Fourier space)
\beq
\gamma_\kp = F_\kp \kpar \vA,
\eeq
where $F_\kp$ is a function of $\kp$ and plasma parameters, but not $\delta z_{\kp}$. Since (refined) critical balance states that (for all structures) $\kpar \vA \sim \tck^{-1}$, the damping factor is
\begin{align}
\gamma_\kp\tck = F_\kp
\longleftrightarrow
\gamma_\lambda \tclam =F_\lambda
\end{align}
where $F_\lambda$ is a function of $\lambda$ but not of $\delta z_\lambda$. This result is true for any intermittency model that incorporates refined critical balance, not solely in the MS17 model \citep{ms16}; it is also the case in the CSM15 model \citep{Chandran14}. Eq.~(\ref{eq:masterdamp}) yields:
\beq
\log\delta z_{\rho-}=\log\delta z_{\rho+}-\frac{2}{3}\gamma_\rho\tcr.\label{eq:gr}
\eeq
Because $\gamma_\rho\tcr$ is independent of $\delta z_{\rho+}$, the effect of the damping is to shift the whole distribution of log-amplitudes over by the constant $(2/3)\gamma_\rho\tcr$; i.e. the \emph{shape} of the distribution is not changed. 
As a corollary, the kurtosis
\beq
\kappa_{\rho-}^{\rm LD} = \frac{\langle \delta z_{\rho-}^4\rangle}{\langle \delta z_{\rho-}^2\rangle^2} = \frac{\langle \delta z_{\rho+}^4\rangle e^{-\frac{8}{3}F_\rho}}{\langle \delta z_{\rho+}^2\rangle^2e^{-\frac{8}{3}F_\rho}} = \kappa_{\rho+}\label{eq:kurtlin}
\eeq
is unchanged. Similarly, the average heating rate per unit volume,
\begin{align}
\langle Q^{\rm LD}_\rho\rangle = \langle\epsilon_{\rho+} - \epsilon_{\rho-}\rangle
=& \left(1-e^{-2F_\rho}\right)\epsilon,\label{eq:qld}
\end{align}
is not affected by the intermittency at all. However, if one looks at \emph{the structures in which the heating is happening}, the intermittency is relevant: The heating rate random variable for each structure,
\beq
Q^{\rm LD}_\rho = \epsilon_{\rho+} - \epsilon_{\rho-} = \left(1-e^{-2F_\rho}\right)\epsilon_{\rho+},
\eeq
follows the (intermittent) distribution of the random variable $\epsilon_{\rho+}$, and damping is concentrated in the higher-amplitude, intermittent structures. Thus, 
Landau damping certainly does not lead to homogeneous wave damping -- a point also made recently by \citet{howes2018}. These results would also apply generically to any damping mechanism for which the damping factor $\gamma_{\rho}\tau_{\mathrm{c}\rho}$ is independent of $\delta z_{\rho}$.

It might be na\"ively thought that these results (Eqs. \ref{eq:kurtlin}, \ref{eq:qld}) are obvious due to the linear nature of Landau damping. Thinking more carefully, these results only apply if the turbulence is critically balanced in the refined sense. For example, if the turbulence were isotropic ($\lpar \sim \lambda$
) at the gyroscale, $\gamma_\rho\tcr \propto 1/\delta z_\rho$. Likewise, if the turbulence were weak ($\lpar\sim \text{const.}$ and $\tc \sim \lambda^2\vA / \lpar \delta z_\rho^2$), $\gamma_\rho\tcr \propto 1/\delta z_\rho^2$. In both these non-critically-balanced cases, the Landau damping is less important in higher-amplitude structures, i.e. the heating rate is more homogeneous than the distribution of $\epsilon_{\rho+}$. This is yet another argument for why critical balance is a crucial organizing principle for magnetized plasma turbulence, and for why one cannot neglect either linear or nonlinear physical phenomena when modelling such turbulence.
\section{Stochastic heating}
The damping rate of gyroscale fluctuations by stochastic heating may be written \citep{chandran2010}\footnote{The damping rate (\ref{eq:sh}) is only expected to be valid at low $\beta_i$; a different formula applies at high $\beta_i$ \citep{hoppock2018}; we do not consider this case.}
\begin{align}
\gamma_{\rho} =& \frac{c_1}{2} \frac{\delta z_\rho}{\rho} \exp\left(-\frac{c_2 v_{\rm th}}{\delta z_\rho}\right).\label{eq:sh}
\end{align}
We take $c_1=0.75$ and $c_2=0.34$ \citep[cf.][]{chandran2010}. The exponential suppression depends on the random variable 
\beq
\xi \equiv \frac{\delta z_\rho}{v_{\rm th}} \sim \beta_i^{-1/2} \frac{\delta z_{\rho+}}{\vA},\label{eq:stcont}
\eeq
where $\beta_i = 8\pi n_i T_i / B_0^2$. Using Eqs.~(\ref{eq:times}) and (\ref{eq:theta}), the damping factor is
\beq
\gamma_{\rho}\tcr = \frac{c_1}{2}\left(\frac{L_\perp}{\rho}\right)^{1/2}\left(\frac{\delta z_{\rho+}}{\delta z_{L_\perp}}\right)\exp\left(-\frac{c_2 \beta_i^{1/2}\vA}{\delta z_{\rho+}}\right),\label{eq:stht}
\eeq
a (highly nonlinear) function of $\delta z_{\rho+}$ \footnote{This is also true using the \citet{Chandran14} intermittency model, although the precise amplitude dependence is different.}. In a qualitative sense, our results on the efficiency of intermittent stochastic heating and its effect on intermittency also apply generically to all mechanisms for which $\gamma_\rho\tau_{\mathrm{c}\rho}$ is an increasing function of $\delta z_{\rho+}$.

\begin{figure}
\includegraphics[width=\linewidth]{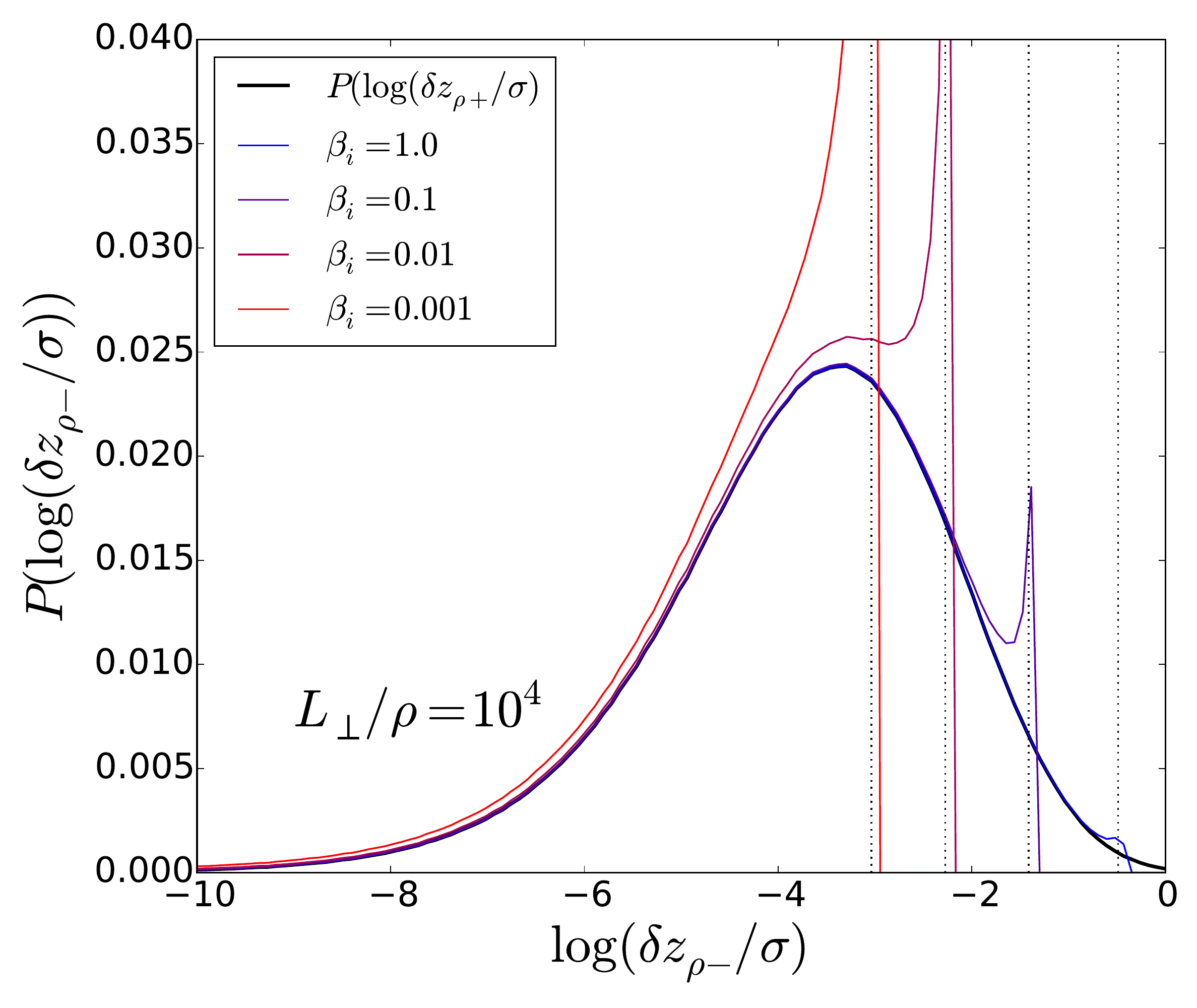}\\
\caption{The distribution of $\log(\delta z_{\rho-}/\sigma)$ resulting from (nonlinear) stochastic heating at $\rho=10^{-4}L_\perp$, for various values of $\beta_i=1,0.1,0.01,0.001$ (blue to red). Vertical dotted lines show $\delta z_{\rm max}$ for each $\beta_i$. The distribution of $\log(\delta z_{\rho+}/\sigma)$ is shown in black. \label{fig:logdists}}
\end{figure}

To illustrate our results, we use a numerically sampled log-Poisson distribution. We take the outer scale amplitudes $\delta z_{L_\perp}$ to be distributed as the magnitude of a normal random variable with zero mean and standard deviation $\sigma=0.1\vA$. We multiply $\delta z_{L_\perp}$ by the log-Poisson factor $\Delta^q$ [Eq.~(\ref{eq:logpois})], generating $10^7$ samples of the intermittent distribution $\delta z_{\rho+}$ just above the gyroscale $\rho$. We then apply damping using Eqs.~(\ref{eq:masterdamp}) and (\ref{eq:stht}) with various different values of $\beta_i$ and $L_\perp/\rho$, obtaining the distributions of $\delta z_{\rho-}$ used in Figures \ref{fig:logdists}--\ref{fig:rhohat_kurt}.
\section{Distribution of fluctuation amplitudes}
The shape of the distribution of $\log(\delta z_{\rho-}/\sigma)$ resulting from stochastic heating is shown for $L_\perp/\rho=10^{4}$ (a value similar to that in the solar wind) and various values of $\beta_i$ (i.e. various different overall damping rates) in Figure \ref{fig:logdists}. As the damping becomes more important (i.e., at lower $\beta_i$), the fluctuations with higher amplitude are heavily damped, causing a relatively sharp upper limit on $\delta z_{\rho-}$. This limit is the amplitude $\delta z_{\rm max}$ for which
\begin{align}
\left.\frac{d\log(\delta z_{\rho-})}{d\log(\delta z_{\rho+})}\right\vert_{\delta z_{\rm max}}=0,\label{eq:qmin}
\end{align}
shown in Figure~\ref{fig:logdists} as a vertical dotted line for each $\beta_i$.

\begin{figure*}
\centering
\includegraphics[width=0.9\linewidth]{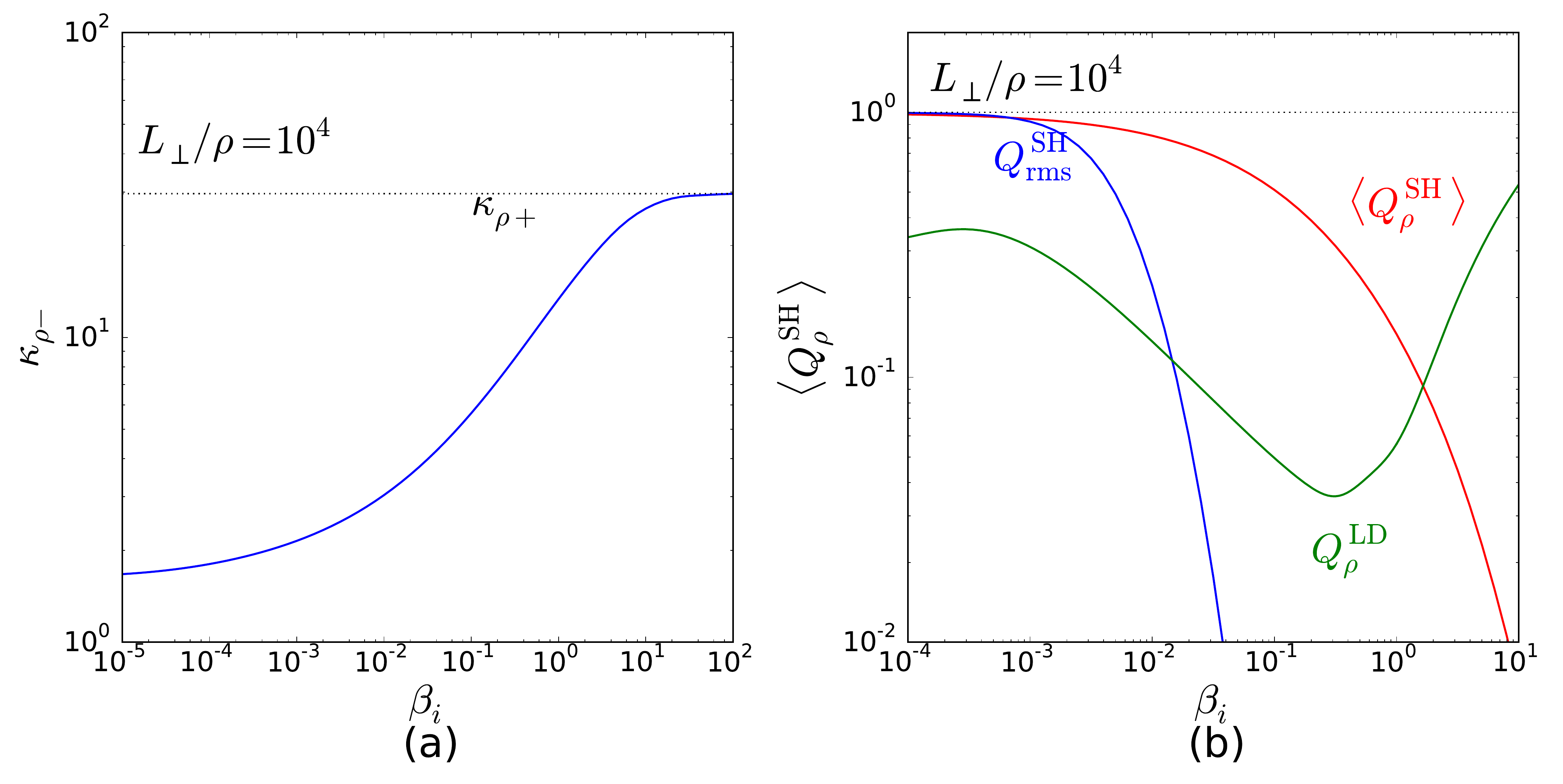}
\vskip-0.4cm
\caption{\textbf{(a)} In blue, the kurtosis after damping, $\kappa^{\rm SH}_{\rho-}$, as a function of $\beta_i$. The black dotted line is the kurtosis without damping $\kappa_{\rho+}$. \textbf{(b)} The heating rates as functions of $\beta_i$: $\langle Q^{\rm SH}_\rho\rangle$ calculated using the intermittent distribution (red), $Q^{\rm SH}_{\rm rms}$ using the rms turbulent amplitude (blue), and $Q^{\rm LD}_{\rho}$, the linear Landau-damping heating rate, normalised to the cascade power $\epsilon$. In both panels, $L_\perp/\rho=10^{4}$ and the outer-scale turbulence amplitude distribution is fixed, parametrised by $\sigma=0.1\vA$ (see text).\label{fig:kurtheat}}
\end{figure*}

Because of this modification of the shape, the kurtosis [Eq.~(\ref{eq:kurt})]
is heavily affected by the damping. In the inertial range, the kurtosis increases as $\lambda$ decreases, reaching a value of $\kappa_{\rho+}= 30$ just above $\rho=10^{-4}L_\perp$. As the stochastic heating becomes more important (with decreasing $\beta_i$), the kurtosis just below $\rho$, $\kappa^{\rm SH}_{\rho-}$, decreases significantly -- see Figure \ref{fig:kurtheat}(a). Such a decrease in kurtosis is a generic property of nonlinear damping mechanisms for which $\gamma_\lambda\tclam$ is an increasing function of $\delta z_\lambda$.
\section{Heating}
Unlike in the linear case, the average stochastic heating rate 
\beq
\langle Q^{\rm SH}_\rho\rangle = \langle\epsilon_{\rho+} - \epsilon_\rho-\rangle = \langle(1-\exp(-2\gamma_\rho\tcr)\epsilon_{\rho+}\rangle
\label{eq:qsh}
\eeq
is affected by the intermittency of the turbulence.
This heating rate may be compared with $\langle\epsilon_{\rho+}\rangle=\langle\epsilon_{L_\perp}\rangle=\epsilon$, and also with the heating rate that would be obtained without intermittency, $Q_{\rm rms}^{\rm SH}$, using the root-mean-square (rms) amplitude $\delta z_{\mathrm{rms}\rho+}\sim \sigma(\rho/L_\perp)^{1/4}$ in place of the random variable $\delta z_{\rho+}$.
These intermittent and rms heating rates, calculated using Eq.~(\ref{eq:stht}) and normalised to $\epsilon$, are shown in Figure \ref{fig:kurtheat}(b), again with $L_\perp/\rho = 10^{4}$. The value of $\beta_i$ at which the damping removes approximately half of the cascade power is significantly higher (by about a factor of 20) with intermittency: $\langle Q^{\rm SH}_{\rho} \rangle\gtrsim0.5$ for $\beta_i\lesssim0.1$, while $Q^{\rm SH}_{\rm rms}\gtrsim0.5$ for $\beta_i\lesssim 0.005$. 

Finally, we calculate the kinetic-Alfv\'en-wave damping rates $\gamma^{\rm KAW}_\rho$ and real frequencies $\omega^{\rm KAW}_\rho$ for $k_\perp\rho =1$ and $k_\parallel/\kp = 10^{-3}$, using the PLUME numerical Vlasov-Maxwell linear dispersion solver \citep{klein2015}, thus estimating the average heating rate from linear Landau damping \footnote{Note that in the low beta regime, Landau damping is dominated by the transfer of energy to electrons.}, $\langle Q^{\rm LD}_{\rho}\rangle$ (using Eq.~(\ref{eq:qld}) with $F_\rho = \gamma^{\rm KAW}_\rho/\omega^{\rm KAW}_\rho$), plotted on Figure~\ref{fig:kurtheat}(b). At the gyroscale, intermittent stochastic heating is comparable to linear Landau damping even for $\beta_i=1$.

\section{Length of the inertial range} The level of intermittency at the gyroscale $\rho$ depends on the length of the inertial range $L_\perp/\rho$ (cf. Eq.~\ref{eq:kurt}). $Q_{\rm rms}^{\rm SH}$ is a strongly decreasing function of $L_\perp/\rho$, simply because the rms amplitude $\delta z_{\mathrm{rms}\rho+}\sim \sigma(\rho/L_\perp)^{1/4}$. The intermittent stochastic heating rate $\langle Q_\rho^{\rm SH}\rangle$ has a weaker dependence on $L_\perp/\rho$, because intermittent, high-amplitude fluctuations in the MS17 model resemble discontinuities with (up to) the outer scale amplitude $\delta z_{L_\perp}$. 

\begin{figure}
\includegraphics[width=\linewidth]{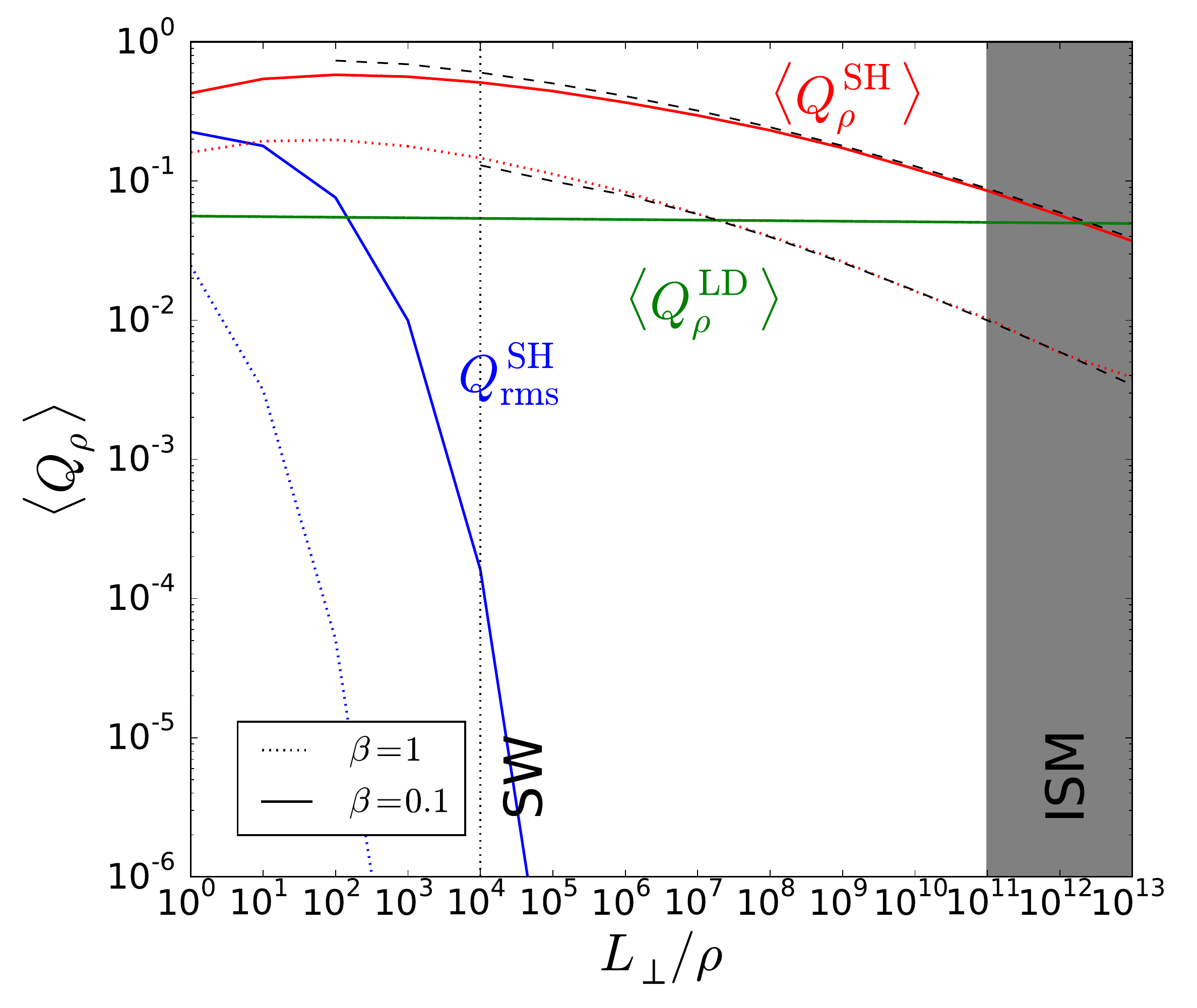}
\caption{Heating rates (normalised to $\epsilon$) $\langle Q^{\rm SH}_\rho\rangle$ (red), $Q_{\rm rms}^{\rm SH}$ (blue), and $Q^{\rm LD}_\rho$ (green), all at $\beta_i=0.1$ (solid lines) and $\beta_i=1$ (dotted lines) as a function of $L_\perp/\rho$. The two curves for $Q^{\rm LD}_{\rho}$ are nearly identical (and thus indistinguishable here -- see Figure \ref{fig:kurtheat}\textbf{(b)}). Approximate ranges of $L_\perp/\rho$ in the solar wind and in the warm ISM are labelled as SW (dotted line) and ISM (gray box). $Q^*_{\rho}$ multiplied by a factor ($4$ for $\beta_i=0.1$ and $6.5$ for $\beta_i=1$) is also shown (black dashed lines).\label{fig:rhohat_kurt}}
\end{figure}

The dependence of the stochastic heating rates for $\beta_i=0.1,1.0$ on $L_\perp/\rho$ are shown in Figure~\ref{fig:rhohat_kurt}, along with $\langle Q^{\rm LD}_\rho\rangle$. The weak dependence of $\langle Q^{\rm SH}_{\rho}\rangle$ on $L_\perp/\rho$ means that, for $\beta_i=0.1$, stochastic heating still removes approximately $10\%$ of the overall cascade power at $L_\perp/\rho \approx 10^{11}$. Moreover, it remains comparable to $\langle Q^{\rm LD}_\rho\rangle$ up to $L_\perp/\rho\approx10^{12}$. Thus, intermittency may have important astrophysical consequences: even at only moderately low $\beta_i$, stochastic heating may (i) convert a large portion of the total cascade power into ion thermal energy at the gyroscale in solar-wind turbulence, where $L_\perp/\rho \approx 10^{4}$, and (ii) be non-negligible (and comparable to linear Landau damping) in the warm interstellar medium (ISM), where $L_\perp/\rho \approx 10^{11}-10^{13}$ \citep{ferriere2001,cox2005,beck2007,haverkorn2008}.

To explain the shallow dependence of $\langle Q_{\rho}^{\rm SH}\rangle$ on $L_\perp/\rho$, we calculate the minimum amplitude $\delta z_{\rho+}^*$ for which fluctuations are strongly damped. Setting $\gamma_{\rho}\tcr = 1$ in Eq.~(\ref{eq:stht}), we obtain
\begin{align}
\log\left(\frac{\delta z_{\rho+}^*}{\delta z_{L_\perp}}\right) = W\left[\frac{c_1c_2\beta_i^{1/2}}{2\sigma}\left(\frac{L_\perp}{\rho}\right)^{\frac{1}{2}}\right]\nonumber\\
-\log\left[\frac{c_1}{2}\left(\frac{L_\perp}{\rho}\right)^{\frac{1}{2}}\right],\label{eq:dzstar}
\end{align}
where $W$ is the Lambert W function. This analytic expression for $\delta z_{\rho+}^*$ approximates $\delta z_{\rm max}$ in Eq.~(\ref{eq:qmin}). If $\delta z^*_{\rho+}$ were determined by simply setting the exponent in Eq.~(\ref{eq:stht}) equal to some constant threshold value, then $\delta z^*_{\rho+}$ would be independent of $L_\perp/\rho$. However, as $L_\perp/\rho$ increases, the fluctuations are increasingly highly aligned [see Eq.~(\ref{eq:theta})] at the gyroscale, which increases $\tcr$ but not $\gamma^{-1}$. This introduces the factor $(L_\perp/\rho)^{1/2}$ in Eq.~(\ref{eq:stht}), causing $\delta z^*_{\rho+}$ to decrease with increasing $L_\perp/\rho$.

The corresponding heating rate from the damping of the structures with this amplitude is
\begin{align}
Q^*_\rho \sim \frac{(\delta z_{\rho+}^*)^2}{\tcr} P(q^*) \sim 
\frac{[\log(L_\perp/\rho) \Delta^2]^{q^*}}{\sqrt{2\pi q^*}(q^*/e)^{q^*}}\epsilon,\label{eq:qstar}
\end{align}
where $q^*=\log(\delta z_{\rho+}^*/\delta z_{L_\perp})/\log\Delta$ [cf. Eq.~(\ref{eq:logpois})], and we have used Stirling's formula to approximate the factorial in the Poisson probability mass function. $Q_{\rho}^*$ is a reasonable analytic estimate for the scaling dependence of $\langle Q^{\rm SH}_{\rho}\rangle$ on $L_\perp/\rho$ for $\log(L_\perp/\rho)\gg1$; however, it is an underestimate (by a factor approximately independent of $L_\perp/\rho$), due to (i) $\delta z^*_{\rho+}$ being an overestimate of the true cutoff, $\delta z_{\rm max}$, (ii) the neglect of the cascade power damped in structures with higher amplitudes $\delta z^*_{\rho+}<\delta z_{\rho+}<\delta z_{L_\perp}$, (iii) the neglect of the width of the outer-scale (normal) distribution of fluctuation amplitudes. For each $\beta_i$, $Q^*_{\rho}$ multiplied by an empirical correction factor is plotted in Figure~\ref{fig:rhohat_kurt}. The analytic expressions for $Q^*_\rho$ and $\delta z^*_{\rho+}$ make clear that the slowly-decreasing nature of $\langle Q_{\rho}^{\rm SH}\rangle$ with $L_\perp/\rho$ arises due to a competition between the decreasing volume-filling fraction of structures above any particular amplitude and the decreasing cutoff amplitude $\delta z_{\rm max}$ ($\approx\delta z_{\rho+}^*$).

In many astrophysical plasmas, $\delta z_{L_0}\sim \vA$ at an outer scale $L_0$ that is beyond the RMHD regime. We can apply our model in such cases on scales $\lambda$ smaller than an effective outer scale $L_\perp\ll L_0$, where $\delta z_{\mathrm{rms}L_\perp} \approx 0.1\vA$. For example, if $\delta z_{\mathrm{rms}\lambda}\propto \lambda^{1/4}$ at $L_\perp<\lambda<L_0$, then $L_\perp = 10^{-4} L_0$. 
The stochastic heating rate when the outer-scale amplitude $\delta z_{L_0} \sim \vA$ is then much larger than our numerical example, where $\delta z_{L_0} \sim 0.1\vA$, because the gyroscale fluctuation amplitudes are much larger. Our figures thus provide a highly conservative lower limit on the stochastic heating rate in plasmas in which $\delta z_{L_0} \sim \vA$.


\section{Conclusions}
In this study, we consider the collisionless damping of critically-balanced, intermittent plasma turbulence by two mechanisms. We develop a new general modelling framework for studying the effects of intermittency on dissipation (and vice versa), and use it to make specific predictions for the efficiency of two different mechanisms. First, for linear Landau damping \citep{landau1946,howes2006,howes2008,howesweakened2011}, $\gamma_\lambda\tclam$ is independent of the turbulent amplitude $\delta z_\lambda$. In this case, (i) damping affects neither the shape of the distribution of log-fluctuation-amplitudes, nor the kurtosis of the distribution of fluctuation amplitudes, and (ii) the overall efficiency of damping is not enhanced by the presence of intermittency. However, (iii) locally, damping is still concentrated near coherent structures \citep{tenbarge2013,howes2018}. Importantly, these results are not an inevitable consequence of the ``linear" nature of linear Landau damping: the efficiency would be strongly affected by intermittency if the turbulence did not obey the refined critical balance \citep{rcb}.

On the other hand, for stochastic heating \citep{chandran2010}, $\gamma_\lambda\tclam$ depends on $\delta z_\lambda$, leading to damping that (i) strongly affects the shape of the distribution of log-fluctuation-amplitudes and the kurtosis of the distribution of fluctuation amplitudes. In addition, stochastic heating is (ii) much more efficient if one accounts for intermittency and (iii) even more concentrated near coherent structures than heating by linear Landau damping. Our results suggest that, once intermittency is incorporated, stochastic heating may be an important damping mechanism for solar-wind turbulence, and perhaps also for some regimes of interstellar turbulence, even when $\xi \sim \delta z_{\rho{\rm rms}}/v_{\rm th}\ll1$ (in which case one would be justified in ignoring stochastic heating if the turbulence were not intermittent). 

Our results can be easily extended to other dissipation mechanisms, which may be divided into different classes based on the (in)dependence of $\gamma_\lambda\tclam$ on $\delta z_\lambda$. This will allow us to quantitatively distinguish between different dissipation mechanisms in observations and simulations of collisionless plasma turbulence. We predict that a nonlinear heating mechanism (for which $\gamma_\lambda\tclam$ is an increasing function of $\delta z_\lambda$) decreases the scale-dependent kurtosis just below the dissipation scale. This leads to a simple observational test to establish the presence of a nonlinear mechanism. 
Indeed, there are numerous observations of decreases in or flattening of the scale-dependent kurtosis at around the ion scale in both numerical and solar-wind turbulence \citep{sundkvist2007,alexandrova2008b,wan2012,wu2013,leonardis2016}; our model provides a natural explanation for this phenomenon (however, we cannot explain why the scale-dependent kurtosis remains rather constant in the range of scales between the ion and electron gyroradii, as in the results of \citet{wu2013} and \citet{chen2014}). Moreover, there is direct evidence for a nonlinear ion heating mechanism, whose efficiency depends on $\xi$ (suggestive of stochastic heating, cf. Eq.~\ref{eq:stcont}), in some numerical simulations \citep{matthaeus2016,groselj2017,shay2018}, while electron heating appears to have $\gamma_\rho\tcr$ independent of $\delta z_\rho$ (suggestive of linear Landau damping; see also \citet{navarro2016} and \citet{chen2019}). Our new modelling framework provides a useful way to interpret these simulation results.

Our results clarify the role of intermittency in heating by collisionless plasma turbulence: since heating rates for nonlinear mechanisms (e.g. stochastic heating) are dramatically enhanced by intermittency, an understanding of the intermittency is essential for determining relative heating rates of different mechanisms, and thus for explaining the eventual thermodynamic state of a turbulent collisionless plasma.
\begin{acknowledgements}
We thank A. A. Schekochihin and R. Meyrand for useful discussions. A. Mallet was supported by by NSF grant AGS-1624501. K.G. Klein was supported by NASA grant NNX16AM23G. C.S. Salem was supported by NASA grant NNX16AI59G and NSF SHINE 1622498. Work at Berkeley is also supported by NASA grants NNX14AJ70G and NNX16AP95G. B.D.G Chandran was supported by NASA grants NNX15AI80, NNX16AG81G, NNN06AA01C, and NNX17AI18G, and NSF grant PHY-1500041.
\end{acknowledgements}
\bibliographystyle{jpp}
\bibliography{mainbib2}
\end{document}